# Mesoporous silica nanoparticles containing silver as novel antimycobacterial agents against *Mycobacterium tuberculosis*

Sandra Montalvo-Quirós[a,b], Sergio Gómez-Graña[c], María Vallet-Regí[c,d], Rafael C. Prados-Rosales[e], Blanca González[c,d,*], Jose L. Luque-García[a,*]

[a] Departament of Analytical Chemistry, Faculty of Chemistry, Complutense University of Madrid. Av. Complutense s/n, 28040, Madrid, Spain.

[b] Centro de Estudios Tecnológicos y Sociales y Facultad de Experimentales, Universidad Francisco de Vitoria, 28223 Pozuelo de Alarcón Madrid, Spain.

[c] Departament of Chemistry in Pharmaceutical Sciences, Faculty of Pharmacy, Complutense University of Madrid, Instituto de Investigación Sanitaria Hospital 12 de Octubre (imas12), Plaza Ramón y Cajal s/n, 28040, Madrid, Spain.

[d] Centro de Investigación Biomédica en Red de Bioingeniería, Biomateriales y Nanomedicina (CIBER-BBN), Spain.

[e] Department of Preventive Medicine and Public Health and Microbiology, Faculty of Medicine, Autonoma University of Madrid.

Sergio Gómez-Graña, Present address: Department of Physical Chemistry and Biomedical Research Center (CINBIO), University of Vigo, Lagoas-Marcosende, 36310 Vigo, Spain

*Corresponding authors*:
jlluque@ucm.es (J.L. Luque-Garcia), blancaortiz@ucm.es (B. González)

Number of words: 5074; 6633 (including references)

Number of Figures: 6

Number of tables: 2

**ABSTRACT**

Tuberculosis remains today a major public health issue with a total of 9 million new cases and 2 million deaths annually. The lack of an effective vaccine and the increasing emergence of new strains of *Mycobacterium tuberculosis* (*Mtb*) highly resistant to antibiotics, anticipate a complicated scenario in the near future. The use of nanoparticles features as an alternative to antibiotics in tackling this problem due to their potential effectiveness in resistant bacterial strains. In this context, silver nanoparticles have demonstrated high bactericidal efficacy, although their use is limited by their relatively high toxicity, which calls for the design of nanocarriers that allow silver based nanoparticles to be safely delivered to the target cells or tissues. In this work mesoporous silica nanoparticles are used as carriers of silver based nanoparticles as antimycobacterial agent against *Mtb*. Two different synthetic approaches have been used to afford, on the one hand, a 2D hexagonal mesoporous silica nanosystem which contains silver bromide nanoparticles distributed all through the silica network and, on the other hand, a core@shell nanosystem with metallic silver nanoparticles as core and mesoporous silica shell in a radial mesoporous rearrangement. Both materials have demonstrated good antimycobacterial capacity in *in vitro* test using *Mtb*, being lower the minimum inhibitory concentration for the nanosystem which contains silver bromide. Therefore, the interaction of this material with the mycobacterial cell has been studied by cryo-electron microscopy, establishing a direct connection between the antimycobactericidal effect observed and the damage induced in the cell envelope.

# 1. Introduction

*Mycobacterium tuberculosis* (*Mtb*) infection affects one third of the world's population, about two billion people, making the current control of tuberculosis an extremely difficult task. It is estimated that 10 million cases of tuberculosis (TB) appear every year and that about 1.5 million people die as a direct consequence of the disease [1]. The World Health Organization has described the increasing microbial resistance to antibiotics as one of the main risks for human health. It has also anticipated the entry of a post-antibiotic era in which common infections could lead to death. This problem is exacerbated by the fact that no new type of antibiotic has been marketed in the last 25 years to which pathogenic microorganisms have not developed resistance. In the case of TB, of the 10 million infections per year, an estimated 500.000 are infections with antibiotic-resistant strains of *Mtb*; in fact, about 80 countries have reported cases of infection with extremely antibiotic-resistant strains [2]. This implies that existing therapeutic strategies for TB will be less effective in the coming years. In this sense, nanomaterials are currently proving to be a promising alternative as antimicrobial therapeutics including *Mtb* infection treatment [3-5].

The antibacterial effect of silver-based nanomaterials has been widely evaluated and described in several studies [6-9]. Their effects are based on the disturbance of bacterial metabolic processes [10-12], the interaction with the DNA of the bacteria [13] or the increase of the cytoplasmic membrane permeability [14,15]. Due to this range of effects, these silver-based nanosystems are effective on both sensitive and highly resistant strains [16,17]. It is important to emphasize that their antibacterial effects, as well as other properties, depend on the particle morphology and its surface characteristics [18,19].

Some authors argue that the main antibacterial mechanism of silver-based nanosystems is due to progressive oxidation to ionic silver ($Ag^+$), while others propose a synergistic effect between the direct interaction of the nanomaterials with the bacterial cell and the released of $Ag^+$ ions [8,20]. Previous studies have tested the activity of several nanomaterials, specifically Ag and AgBr nanoparticles (AgNPs and AgBrNPs), in different types of bacteria [21]. In general, it has been demonstrated that there are differences in antibacterial activity that arise from the differential interaction of silver-based materials with the cell walls of microorganisms. In the case of Gram-positive bacteria, it has been observed greater sensitivity to AgNPs, probably due to the destructive interaction of these materials with cell walls. However, Gram-negative bacteria and yeast strains exhibited greater sensitivity to AgBrNPs due to the combination of the direct interaction of the NPs with the cell walls and the simultaneous release of $Ag^+$ able to penetrate into the cells [21]. These differences shown in the antibacterial activity as a function of the nanomaterial used may provide a basis for the development of selective targeted antibacterial nanosystems for a specific strain type. Also, Ag/AgBr nanostructures have been reported with an enhanced antibacterial activity due to photodynamic effects and the innate antimicrobial ability of $Ag^+$ [22]. Furthermore, AgNPs alone or in combination with biomolecules have good antimycobacterial effects and are promising future therapeutics for TB because overcome the current antibiotic resistance [23].

Despite the bactericidal potential of AgNPs and AgBrNPs, it is necessary to take into account their high toxicity [24-26], which makes their use in biomedical applications limited. Therefore, it is required their carrying in another vehicle to allow a more selective administration, as in the case of cancer-cell targeting [27,28]. This strategy avoids systemic exposure to metal nanoparticles, as well as prevents the aggregation problems associated to uncoated AgNPs and AgBrNPs [29]. In this context, the use of mesoporous silica nanoparticles (MSNs) as carriers has attracted the nanomedicine scientists due to their biocompatibility and unique properties, such as particle diameters in the 50–200 nm range, highly ordered pore network with narrow pore size distributions of 3–6 nm and silanol rich surface that allows their functionalization. MSNs represent a versatile nanoplatform able to be combined with organic and inorganic moieties for biomedical applications, such as cancer and infection treatment [30-36]. Aimed at TB treatment, MSNs have been reported as delivery vehicles of antimicrobial agents against mycobacterium *Mtb* such as the antimycobacterial peptide NZX [37], the water insoluble antibiotic clofazimine [38] or the antibiotic isoniazid [39-41]. The major form of TB disease is manifested in the pulmonary form and MSNs have been also investigated for aerosol delivery representing an efficient strategy for respiratory-based therapeutics [42].

However, to the best of our knowledge, MSNs as carriers of silver with antimycobacterial activity against *Mtb* have not yet been reported. MSNs containing silver would overcome the aggregation problems of AgNPs and would present advantages such as the possibility of functionalization to increase local availability, even macrophage internalization, and the combination with other anti-TB drugs allocated into the mesopores for a synergic antimycotic activity. Based on these premises, in this work two different mesoporous silica-silver hybrid nanosystems with potential antimycobacterial effect against *Mtb* have been designed and tested. While one of the nanosystems consists of AgBr nanoparticles supported on the MSNs network, the other consists of MSNs containing an Ag core in a core@shell configuration. After nanosystems synthesis and characterization, their antimycobacterial capacity of has been assessed by measuring their minimal inhibitory capacity. The AgBrNPs based nanosystem shows the best inhibition of bacterial growth and, therefore, it has been used to analyze its effect on the cell envelope of *Mtb* using cryo-EM.

## 2. Experimental

*2.1. Chemicals*

Tetraethylorthosilicate (TEOS), cetyltrimethylammonium bromide (CTAB), silver nitrate 99.9% and sodium borohydride were purchased for Sigma-Aldrich. *N*-(amino-ethyl)-amino-propyl trimethoxysilane (TSD) was purchased from ABCR GmbH y Co.KG. The resazurin reagent was obtained as resazurin sodium salt powder (Sigma-Aldrich). All other chemicals (absolute EtOH, NaOH, etc.) were of the highest quality commercially available and used as received. Milli-Q water (resistivity 18.2 MΩ·cm at 25 °C) was used in all experiments.

*2.2. Synthesis of MSNs-AgBrNPs*

A one-pot route for the incorporation on silver ions followed by a thermal treatment was followed for the incorporation of silver bromide in the MSNs framework [43]. CTAB (0.3 g, 8.23 mmol) was dissolved in 110 mL of water and 1 mL of NaOH 2 M was added. The mixture was stirred and heated at 75 °C. TEOS (2 mL, 9.02 mmol) was dissolved in 10 mL of ethanol, and 8 mL of the resulting solution was added dropwise at a constant rate (0.8 mL/min) over the previous CTAB solution. The mixture was reacted for 15 min under vigorous stirring to form mesoporous silica cores. $AgNO_3$ (0.06 g, 3.53 mmol) was dissolved in 10 mL of water and mixed with TSD (0.5 mL, 2.27 mmol) under stirring. The resulting solution was added dropwise at a constant rate (1.5 mL/min) over the solution containing TEOS and CTAB. Then, the remaining solution of TEOS was also added at constant rate of 0.8 mL/min. The reaction was stirred at 75 °C for 2 h. Then, nanoparticles were collected by centrifugation and washed with water and ethanol. Finally, the solid was subjected to a thermal treatment in a tubular oven by heating at 10 °C/min to reach 600 °C and this final temperature was maintained for 3 h under a constant rate of air stream.

*2.3. Synthesis of Ag@MSNs*

40 mL of 0.1 M CTAB were mixed with 10 mL of 0.1 M $AgNO_3$ in an opaque flask of 100 mL at 30 °C to avoid precipitation of CTAB. Then, 10 mL of 0.2 M $NaBH_4$ were added under vigorous stirring. The resulting solution containing silver nanoparticles (AgNPs) was kept overnight under stirring at 30 °C in the absence of light. The mesoporous silica layer around the AgNPs was formed following the protocol previously reported by Sanz-Ortiz *et al.* [44] by using AgNPs instead of AuNPs. At 35 °C and under magnetic stirring, 170 mL of 6 mM CTAB were sequentially mixed with 70 mL of ethanol, 100 µL of $NH_3$, 4 mL of the freshly prepared AgNPs and 200 µL of TEOS. When TEOS was added, the temperature was increased up to 50 °C. The reaction was kept in the darkness overnight. The obtained solid was washed by centrifugation at 14000 rpm for 30 min to remove the excess of reactants. Extraction of the surfactant (CTAB) from the as-prepared nanomaterial, was performed by dispersing Ag@MSNs-CTAB in 200 mL of an extracting solution containing: 1.59 g of $NH_4NO_3$ in 573 mL of absolute ethanol and 27 mL of water. The mixture was heated at 70 °C and stirred overnight. Then, the Ag@MSNs solution was washed by centrifuging and kept dispersed in ethanol. Surfactant removal was confirmed by thermogravimetry and infrared spectroscopy.

*2.4 Analytical characterization of the nanomaterials*

The analytical techniques used for the characterization of the synthesized materials were as follows: thermogravimetric and differential thermal analysis (TGA and DTA), Fourier transformed infrared (FTIR) spectroscopy, UV-visible spectroscopy, low- and high-angle powder X-ray diffraction (XRD), inductively coupled plasma atomic emission spectroscopy (ICP-AES), transmission electron microscopy (TEM), energy dispersive X-ray spectroscopy (EDX), $N_2$ sorption porosimetry, electrophoretic mobility measurements to calculate the values of the zeta-potential ($\zeta$) and dynamic light scattering (DLS). The equipment and conditions used are described in the Supplementary Material. For the silver and silicon release test, 5 mg of each material was suspended in 1.5 mL of Middlebrook 7H9 medium and maintained at 37 °C under orbital shaking for the different times assayed. For each time, the supernatant was recovered by centrifuging and replaced by fresh medium.

*2.5. Mycobacteria culture conditions*

*M. tuberculosis* H37Rv strain (*Mtb*) obtained from the ATCC was grown in Middlebrook 7H9 medium supplemented with 10% (v/v) OADC supplement (NaCl 8.5 g/L, BSA fraction V 50 g/L, dextrose 20 g/L, 5% (v/v) oleic acid solution 1%, 40 mg/L catalase, 0.5% (v/v) glycerol and 0.05% Tyloxapol (v/v). Cultures were grown at 37 °C in static standing 25 $cm^2$ flasks with vented caps.

*2.6. Minimum inhibitory concentration assay*

Minimum inhibitory concentration (MIC) assay was carried out in 96-well plates. *Mtb* was cultured at an initial density of $1 \times 10^5$ bacteria/mL in the presence of the indicated concentrations of nanoparticles (MSNs-AgBrNPs and Ag@MSNs). Plates were incubated for 20 days at 37 °C. Mycobacterial growth was monitored at day 4, 14 and 20 by measuring optical density at 570 nm. The assay was performed in triplicate. MIC values were selected as the minimum concentration able to suppress mycobacterial growth. Alternatively, MIC was determined for the indicated nanoparticles by using the resazurin method as previously described [45]. Microtiter plates were set up as above and incubated at 37 °C for 5 days. Then, a volume of 100 ul of a sterile 0.01% wt/vol resazurin solution was added to each well and incubated for 24 h at 37 °C. Wells showing a color change from blue (oxidized state) to pink (reduced state) indicated bacterial viability.

*2.7. Cryo-electron microscopy (Cryo-EM)*

*Mtb* cells treated with a sub lethal concentration of MSNs-AgBrNPs were fixed with 2% glutaraldehyde in 0.1 M cacodylate at room temperature for 2 h, and then incubated overnight in 4% formaldehyde, 1% glutaraldehyde, and 0.1% PBS. Grids were prepared following standard procedures and observed at liquid nitrogen temperatures in a JEM-2200FS/CR transmission electron microscope (JEOL Europe, Croissy-sur-Seine, France) operated at 200 kV. An in-column omega energy filter helped to record images with improved signal/noise ratio by zero-loss filtering. The energy selecting slit width was set at 9 eV. Digital images were recorded on an UltraScan4000 CCD camera under low-dose conditions at a magnification of 55,058 obtaining a final pixel size of 2.7 Å/pixel.

## 3. Results and discussion

*3.1. Synthesis of materials*

The synthesis of mesoporous silica nanoparticles containing supported silver bromide nanoparticles (MSNs-AgBrNPs) was carried out in a one-pot route for the incorporation on silver ions followed by a thermal treatment (Figure 1A). Organic-inorganic hybrid mesoporous silica nanoparticles were initially formed by co-condensation of the silica precursor TEOS and the alkoxysilane TSD which contains both primary and secondary amine functional groups as anchorage points for the complexation of $Ag^+$ ions. The co-condensation was performed in the presence of CTAB as structure directing agent. Also, silver nitrate was present during the sol-gel reactions, therefore affording the complexation of $Ag^+$ ions in the amino groups attached to the silica framework. Finally, the thermal treatment affords the removal of the organic surfactant and the formation of AgBrNPs simultaneously in the same step.

The preparation of core@shell nanoparticles (Ag@MSNs) was performed in a two-stage process (Figure 1B). The silver nanoparticles that form the core of the final material were prepared in the first step. For this purpose, the ionic precursor $AgNO_3$ underwent a chemical reduction process with sodium borohydride according to the following reaction:

$$2\ AgNO_3 + 2\ NaBH_4 + 6\ H_2O \rightarrow 2\ Ag + 2\ NaNO_2 + 2\ B(OH)_3 + 7\ H_2O$$

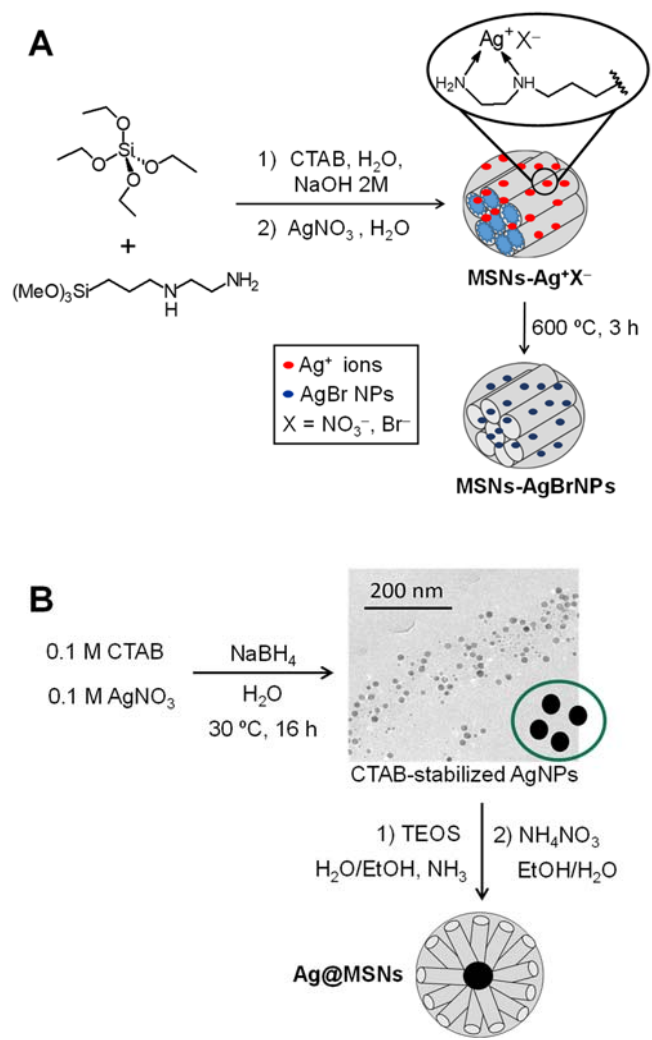

**Figure 1.** Scheme of synthesis of A) mesoporous silica nanoparticles containing silver bromide nanoparticles (MSNs-AgBrNPs) following a one-pot incorporation of silver ions and subsequent thermal treatment and B) core@shell type nanoparticles consisting of silver coated by mesoporous silica (Ag@MSNs).

The obtained AgNPs were characterized by TEM and UV-visible spectroscopy. The TEM micrographs showed a homogeneous dispersion of AgNPs with a diameter of *ca.* 14 nm (Figure S1A and S1D). The UV-visible spectrum (Figure S1C) showed a maximum absorbance peak at 408 nm which corresponds to the plasmon band of silver nanoparticles smaller than 20 nm, thus confirming the diameter observed by TEM [46]. In a second step, the AgNPs stabilized with CTAB were coated via hydrolysis and condensation of TEOS as silica source. The surfactant CTAB acts as stabilizing agent to prevent the silver nanoparticles from aggregation as well as template or structure directing agent during the formation of the mesoporous silica network in the sol-gel synthesis. Micrographs obtained by TEM of the final material showed core@shell nanoparticles with a homogeneous diameter of *ca.* 77 nm (Figure S1B and S1D). The shift of the maximum plasmon peak to 421 nm, as well as the increase of its full width at half maximum, observed in the UV-visible spectra of the Ag@MSNs material (Figure S1C) maybe ascribed to the mesoporous silica coating observed in the TEM study.

*3.2. Materials characterization*

The incorporation of silver in the mesoporous silica nanoparticles was first assessed by TEM. Figures 2A and S2 show micrographs of MSNs-AgBrNPs material, where inorganic nanoparticles of *ca.* 12 ± 6 nm supported in the silica network can be seen and the absence of free silver-based nanoparticles is confirmed. The mesoporous channels are observed in a parallel arrangement similar to the MCM-41 type mesostructure, although somehow distorted due to the distribution of the AgBrNPs throughout the silica. Micrographs of the Ag@MSNs material (Figure 2B and S2) show a core@shell type configuration for the nanoparticles which very homogeneous spherical shape and size distribution of *ca.* 70 nm. The silver nanoparticles constituting the nucleus have homogenous size distribution of *ca.* 10 ± 2 nm and the mesoporous silica shell presents a radial arrangement.

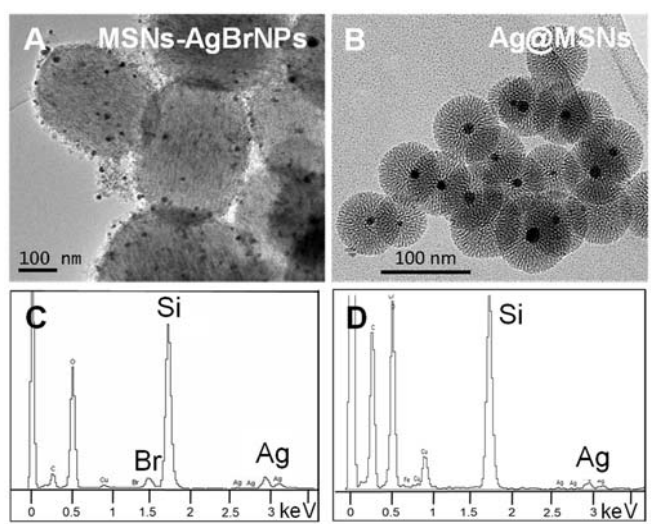

**Figure 2.** TEM micrographs and EDX analysis of MSNs-AgBrNPs (A, C) and Ag@MSNs (B, D) materials, respectively.

The elemental analysis of materials was performed by EDX. The spectrum of Ag@MSNs only presented signals corresponding to Si and Ag (Figure 2D). However, the MSNs-AgBrNPs material also presented a signal corresponding to Br in its EDX spectrum, pointing out to the possibility of AgBr as inorganic nanoparticles in this material (Figure 2C). The atomic percentages are shown in Table S1, and their analysis indicates a molar ratio Ag/Si of 0.0317 for MSNs-AgBrNPs and 0.0342 for Ag@MSNs, thus being comparable the amount of silver incorporated in both silica-based nanosystems. The low content in Br found in the Ag@MSNs material could be due to residues of the surfactant CTAB used in the stabilization of AgNPs and later as template of the mesostructure of the silica since this material has not been subjected to a thermal treatment for the removal of the surfactant.

The nature of the silver based nanoparticles present in the materials was further studied by high angle X-ray diffraction. Figure 3B shows the diffractograms obtained for both materials where a different pattern can be observed for each one. While the Ag@MSNs material exhibits the pattern corresponding to metallic silver with a face-centered cubic structure (JCPDS, No.4-0783), the MSNs-AgBrNPs material shows a pattern indexed to the cubic structure of AgBr (JCPDS, No. 06-0438). XRD maxima, lattice spacing and hkl indexes of diffractograms for each material are listed in Table 1. These results, together with the absence of maxima due to crystalline impurities, confirm the presence of different pure crystalline phases in each nanosystem, $Ag^0$ nanoparticles in the Ag@MSNs material and AgBr nanoparticles in the case of the MSNs-AgBrNPs material.

**Table 1.** XRD maxima and lattice spacing ($d_{hkl}$) of MSNs-AgBrNPs and Ag@MSNs materials.

| Material | 2θ | $d_{hkl}$ | (hkl) |
|---|---|---|---|
| Ag@MSNs | 38.03 | 0.236 | 111 |
| | 44.38 | 0.203 | 200 |
| | 64.45 | 0.144 | 220 |
| | 77.42 | 0.123 | 311 |
| | 81.51 | 0.117 | 222 |
| MSNs-AgBrNPs | 30.93 | 0.288 | 200 |
| | 44.37 | 0.204 | 220 |
| | 55.09 | 0.166 | 222 |
| | 64.59 | 0.144 | 400 |
| | 73.28 | 0.128 | 420 |

Regarding the mesoporous structure, the results obtained in the low angle region of the XRD measurements (Figure 3A and Table S3) agree with the TEM studies. The MSNs-AgBrNPs material presents a pattern referable to the MCM-41 structure with a 2D-hexagonal plane array of pores with *p6mm* symmetry (Figure S3). The diffraction maxima can be indexed as 10, 11 and 20 indexes of a *p6mm* space group, and a unit cell parameter of 4.05 nm can be calculated using the equation $a_0 = (2/\sqrt{3})d_{10}$. The broad maximum corresponding to the 10 reflection may indicate a distortion of the honeycomb mesoporous arrangement typical of MCM-41 structure, probably due to the presence of AgBrNPs distributed throughout the silica network. The diffractogram of Ag@MSNs material shows just one broad maximum around 1.77 degrees, reflecting the radial mesoporous arrangement of the silica shell observed in the TEM micrographs.

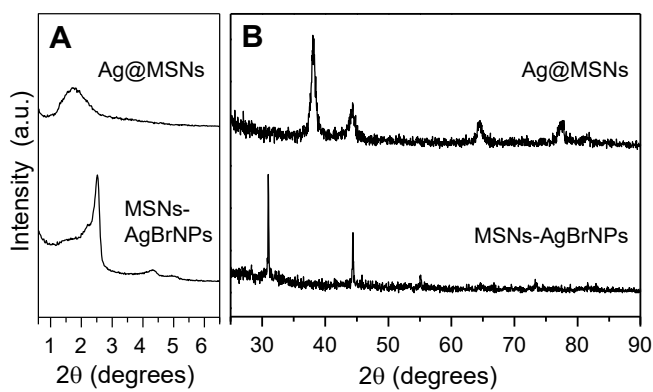

**Figure 3.** Powder X-ray diffraction patterns of the MSNs-AgBrNPs and Ag@MSNs nanosystems registered in the low (A) and high (B) angle regions.

The textural properties of mesoporous silica nanoparticles were studied by $N_2$ sorption analysis (Figure 4 and Table 2). The curves for both materials exhibit a type IV Brunauer–Emmett–Teller (BET) adsorption-desorption isotherms with no observed hysteresis loop, confirming the mesoporous structure of materials exhibiting cylindrical pores. The $N_2$ isotherms of MSNs-AgBrNPs and Ag@MSNs present an inflection at a relative pressure of 0.18–0.30 and 0.20–0.40, respectively, which corresponds to the phenomena of capillary condensation and evaporation within channel-type mesopores. Neither curve shows a very sharp increase, as expected due to the loss of uniformity of the cylindrical channels with respect to a typical MCM-41 type mesoporous structure. The secondary step observed at a pressure above 0.90 $P/P_0$ is attributed to interparticle $N_2$ condensation and the high amount of adsorbed $N_2$ in this region for the Ag@MSNs material reveals the smaller size of the

core@shell nanosystem. Due to this difference in the interparticle condensation, pore volume has been also calculated at P/P$_0$ = 0.60 for comparison of both materials. The nanosystems possess high surface area and pore volume values (Table 2), although are lower for the MSNs-AgBrNPs material, maybe reflecting a partial blocking of the mesopores due to the presence of AgBrNPs in the silica framework. For the core@shell nanosystem, the AgNPs are in the central core thus being the mesopores unrestricted. Pore diameter is around 2.5 nm for both materials, while is higher for the Ag@MSNs material, also consistent with the interplanar distance obtained for the XRD lattice spacing of the (10) reflection (Table S3).

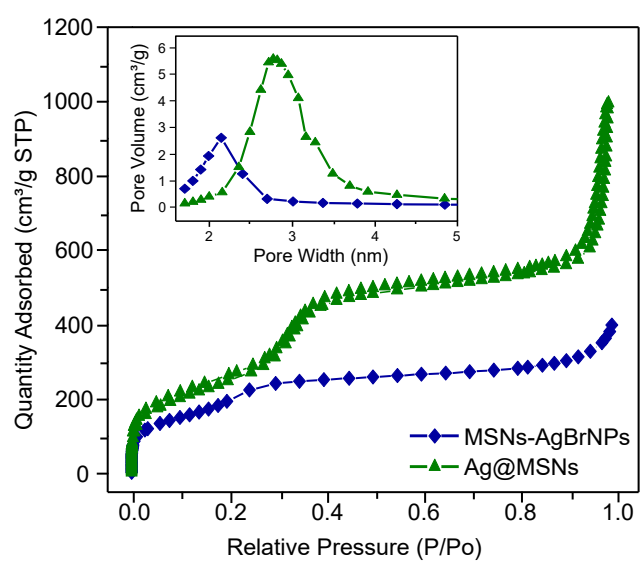

**Figure 4.** N$_2$ sorption isotherms of MSNs-AgBrNPs and Ag@MSNs materials. The inset shows the corresponding pore size distributions for the mesoporous samples.

**Table 2.** Characteristics of the materials synthesized in this work obtained by low-angle XRD, N$_2$ adsorption measurements, ζ-potential and dynamic light scattering.

| Material | 2θ (°) | d$_{10}$ (nm) | S$_{BET}$ (m$^2$/g) | D$_p$ (nm) | V$_T$ (cm$^3$/g) | V$_P$ (cm$^3$/g) | ζ-potential (mV) | Hydrodynamic size (nm) |
|---|---|---|---|---|---|---|---|---|
| MSNs-AgBrNPs | 2.51 | 3.505 | 723.0 | 2.15 | 0.54 | 0.41 | −19 ± 3 | 245.3 ± 4.8 |
| Ag@MSNs | 1.77 | 4.963 | 920.1 | 2.80 | 1.54 | 0.77 | −19 ± 6 | 91.3 ± 1.6 |

d$_{10}$: XRD lattice spacing of the (10) reflection. Textural parameters: S$_{BET}$, specific surface area obtained by using the BET equation; D$_p$, pore diameter calculated by using BJH method; V$_T$, total pore volume obtained at P/P$_0$ = 0.97; V$_P$, total pore volume obtained at P/P$_0$ = 0.60. Hydrodynamic size is the maximum of the size distribution measured by DLS.

The ζ-potential of MSNs-AgBrNPs and Ag@MSNs materials presents a negative value of approximately −19 mV in both cases, high enough to be within the colloidal stability zone (Table 2). This value is due to the presence of negatively charged silanol groups (-Si-O⁻) following the acid-base equilibrium of silanol groups of the silica surface in water:

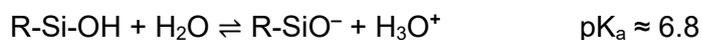

R-Si-OH + $H_2O$ ⇌ R-SiO⁻ + $H_3O^+$        $pK_a \approx 6.8$

The similarity of values for both materials indicates that AgBrNPs incorporation throughout the silica network in MSN-AgBrNPs does not affect the silica surface properties of material, being comparable to the core@shell nanosystem.

The hydrodynamic size was obtained by dynamic light scattering measurements showing monomodal distributions for both MSNs-AgBrNPs and Ag@MSNs, which indicates the presence of only one population of nanoparticles and therefore discarding the occurrence of free silver based NPs (Figure S4). The maximum of the distribution is *ca.* 90 nm for the Ag@MSNs material and *ca.* 245 nm for MSNs-AgBrNPs (Table 2) confirming the smaller size for the Ag@MSNs nanosystem, as above concluded in TEM and $N_2$ adsorption characterization.

*3.3. Antimycobacterial activity of the materials*

In order to evaluate the antimycobacterial properties of the materials against *Mtb*, the minimum inhibitory concentration (MIC) was measured at days 4, 14 and 20 after treatment with serial suspensions of MSNs-AgBrNPs and Ag@MSNs nanosystems, starting from an initial concentration of 500 μg/mL of each material. The results obtained showed an inhibition of bacterial growth by the effect of both materials (Figure 5). The value of MIC of the evaluated materials can be defined as the minimum concentration that inhibits the visible growth of the evaluated microorganism. For all the time points evaluated, MSNs-AgBrNPs exhibited a more effective antimycobacterial effect as compared to Ag@MSNs (Figure 5). After 20 days (Figure 5C), the MIC value for the MSNs-AgBrNPs material was 31.25 μg/mL, while for the Ag@MSNs material it was 250 μg/mL. Further, we found that MSNs-AgBrNPs truly compromise bacterial viability as determined by the resazurin test (Figure S5 in Supplementary Material). Using this test, we could measure that MSNs-AgBrNPs affect bacterial viability at a concentration of 15 μg/mL.

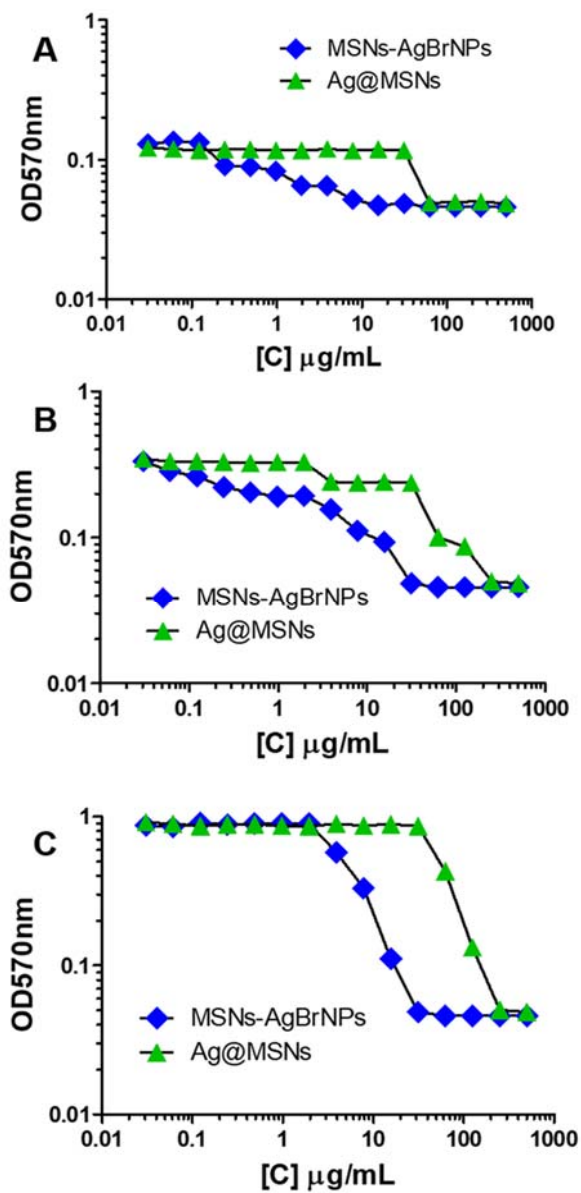

**Figure 5.** MIC assay testing the antimycobacterial activity of MSNs-AgBrNPs and Ag@MSNs against *Mtb* at days 4 (A), 14 (B) and 20 (C).

Since both materials have similar Ag content with respect to Si, the more accentuated mycobactericidal effect against *Mtb* observed for MSNs-AgBrNPs could be explained by the presence of AgBrNPs distributed throughout the silica network and the mesoporous channels and therefore more exposed to the media, compared to the Ag core present in Ag@MSNs. Furthermore, the release of Ag$^+$ ions, capable of penetrating the mycobacterial membrane [21,47] and possibly responsible for the mycobactericidal action, might be faster from AgBrNPs than from Ag cores, where such release depends on the previous oxidation of Ag$^0$ to Ag$^+$ [48]. A silver and silicon release experiment was carried out under the *Mtb* culture conditions to try to justify the different behavior of both materials. The concentration of the elements was determined in the solution at different times up to 28 days by ICP-AES. Silver and bromine were not detected in the release medium, getting values lower than the detection limit of the equipment for all times recorded. However, silicon presented concentrations from 47 to 346 mg/L and 23 to 261 mg/L for MSN-AgBrNPs and Ag@MSNs, respectively (see Table S2 in Supplementary material). On the other hand, the solids recovered after 5 days of immersion in the culture medium were analyzed by TEM and EDS (Figure S6 and Table S1). The TEM images confirm the silica degradation in both samples, as expected for mesoporous silica nanoparticles in aqueous media [49]. Regarding the silver, in the MSN-AgBrNPs material, the AgBr nanoparticles have suffered a reorganization and are found in agglomerates independent from the MSNs themselves. However, the Ag@MSNs material preserves the Ag core surrounded by the degraded mesoporous silica shell. Consistent with these images, the corresponding EDS analyses confirm a six- and seven-fold increase in the Ag/Si molar ratio with respect to the initial materials, due to the fact that the amount of silver is maintained and the silicon is leached into the medium. However, we cannot discard silver release in the antimycobacterial activity assay, although in the *Mtb* culture medium in the absence of the mycobacteria, silver could be oxidizing and reducing not being detected as soluble $Ag^+$ neither short nor long times assayed.

Based on the better effective antimycobacterial effect of MSNs-AgBrNPs, the interaction of this nanosystem with *Mtb* was further analyzed attending to the fact that nanoparticles can compromise the integrity of bacterial membranes [21].

*3.4. Morphological study of Mycobacterium tuberculosis after treatment with MSNs-AgBrNPs*

To gain insight into the nature of the inhibitory effect of MSNs-AgBrNPs, a study in a close-to-native state of the morphological changes in the surface of *Mtb* after treatment with MSNs-AgBrNPs was carried out by cryo-EM. Micrographs of the untreated *Mtb* cells showed an intact morphology with normal cell wall and organization of envelope layers (Figure 6A). However, *Mtb* cells treated with MSNs-AgBrNPs presented membrane damage, as seen in Figure 6B. In addition, nanoparticles can be observed in direct contact with the *Mtb* cell membrane, which suggests the mycobactericidal effect and the damage detected in the bacterial envelope may be also related to the direct interaction between *Mtb* and the nanosystem. These results agree with previous studies which show that nanoparticles can influence the bacterial cell wall by direct contact without the need to be endocytosed [50,51].

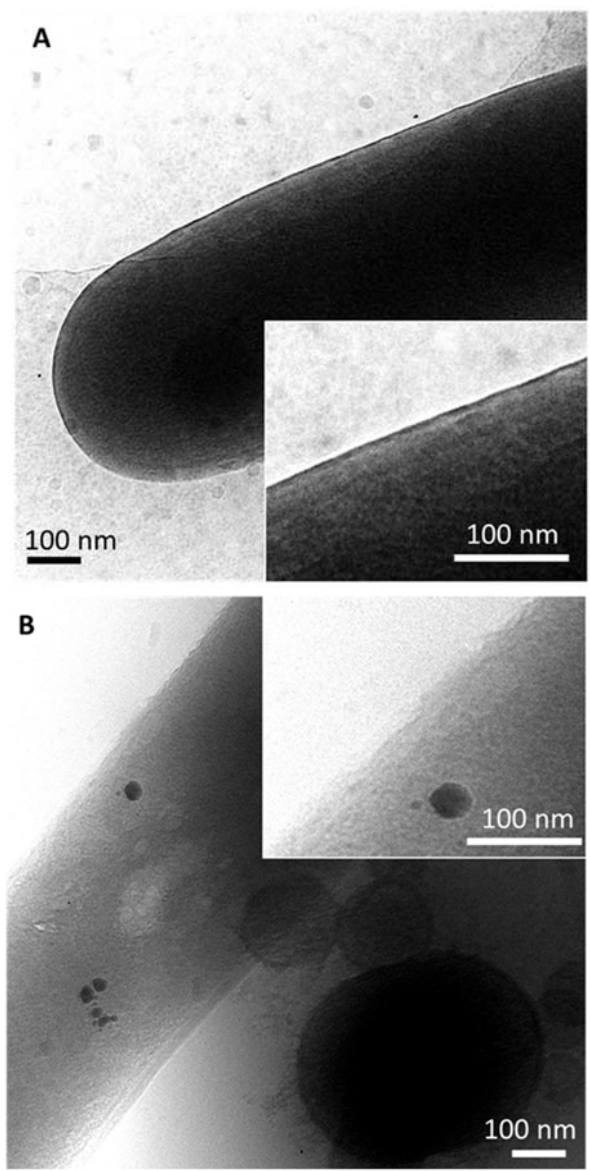

**Figure 6.** Ultrastructural analysis by cryo-EM of untreated *Mtb* cells (A) and *Mtb* cells treated with an inhibitory concentration of MSNs-AgBrNPs (B).

## 4. Conclusions

In the present work two different silver containing mesoporous silica based nanosystems have been prepared as potential antimycobacterial agents against tuberculosis. Physico-chemical characterization confirms the successful synthesis and homogeneity of both nanosystems, ensuring the presence of silver bromide nanoparticles distributed throughout the silica network in the MCM-41 type nanosystem and a metallic silver nanoparticles nucleus in the core@shell configuration of the other nanosystem. *In vitro* antibacterial tests in *Mycobacterium tuberculosis* allows determining the minimum inhibitory concentration for each material. Both nanosystems showed satisfactory results, having the best antimycobacterial properties against *Mtb* the silver bromide containing nanosystem MSNs-AgBrNPs. The microscopic evaluation of the mycobacterial cell wall after treatment with MSNs-AgBrNPs confirms the direct connection between the effect of MSNs-AgBrNPs and the reduction of the integrity of the *Mtb* cell envelope. Overall, the results demonstrate the great potential of these nanomaterials as antimycobacterial agents against *Mtb*. The fact that both nanosystems are based on mesoporous silica nanoparticles represents a great advantage due to the possibility of carrying the silver-based nanoparticles avoiding their aggregation and opening the possibility to selective cell targeting to avoid as well toxicity, thanks to the versatility of functionalizing MSNs with specific ligands. On the other hand, additional drugs such as antibiotics could be accommodated within the mesopores allowing for the design of a multi-drug nanosystem for the selective targeting of resistant tuberculosis strains.

**Declaration of interests**

The authors declare that they have no known competing financial interests or personal relationships that could have appeared to influence the work reported in this paper.

**Acknowledgements**

This study was supported by Ministerio de Ciencia e Innovación grants: CTQ2017-85673-R (JLL-G), SAF2016–77433-R (RP-R) and the European Research Council ERC-2015-AdG (VERDI) Proposal No. 694160. R.P.-R. is further a 'Ramon y Cajal' fellow from the Spanish Ministry of Economy and Competitiveness. S.G.-G. acknowledges the Juan de la Cierva-Formación Program (FJCI-2014-20186). CIBER is a public research consortium created by ISCIII whose actions are co-funded by the European Regional Development Fund.

**Supplementary material**

**Mesoporous silica nanoparticles containing silver as novel antimycobacterial agents against *Mycobacterium tuberculosis***

Sandra Montalvo-Quirós[a,b], Sergio Gómez-Graña[c], María Vallet-Regí[c,d],
Rafael C. Prados-Rosales[e], Blanca González[c,d],*, Jose L. Luque-García[a],*

[a] Departament of Analytical Chemistry, Faculty of Chemistry, Complutense University of Madrid. Av. Complutense s/n, 28040, Madrid, Spain.

[b] Centro de Estudios Tecnológicos y Sociales y Facultad de Experimentales, Universidad Francisco de Vitoria, 28223 Pozuelo de Alarcón Madrid, Spain.

[c] Departament of Chemistry in Pharmaceutical Sciences, Faculty of Pharmacy, Complutense University of Madrid, Instituto de Investigación Sanitaria Hospital 12 de Octubre (imas12), Plaza Ramón y Cajal s/n, 28040, Madrid, Spain.

[d] Centro de Investigación Biomédica en Red de Bioingeniería, Biomateriales y Nanomedicina (CIBER-BBN), Spain.

[e] Department of Preventive Medicine and Public Health and Microbiology, Faculty of Medicine, Autonoma University of Madrid.

*Corresponding authors*:

jlluque@ucm.es (J.L. Luque-Garcia), blancaortiz@ucm.es (B. González)

**Characterization techniques**

Thermogravimetric analysis (TGA) and differential thermal analysis (DTA) were performed in a Perkin Elmer Pyris Diamond TG/DTA analyser (Perkin Elmer, California, USA) by placing approximately 5 mg of sample in an aluminum crucible applying 5 °C/min heating ramps from room temperature to 600 °C under a flow rate of 100 mL/min of air.

Fourier transformed infrared (FTIR) spectra were collected in a Thermo Nicolet Nexus spectrometer equipped with a Goldengate attenuated total reflectance (ATR) device.

UV-visible spectroscopy measurements were performed in a HELIOS-ZETA spectrophotometer.

Inductively Coupled Plasma Atomic Emission Spectroscopy (ICP-AES) was performed in a Varian Vista AX Pro spectrometer through the Si, Ag and Br emission lines at 250.690, 328.068 and 780,302 nm, respectively.

Powder X-ray diffraction (XRD) measurements were performed in a Philips X'Pert diffractometer with Bragg-Brentano geometry operating with Cu K$\alpha$ radiation (wavelength 1.5406 Å) at 40 kV and 20 mA (Philips Electronics NV, Eindhoven, Netherlands). Low-angle XRD patterns were collected in the 2$\theta$ range between 0.6° and 8° with a step size of 0.02° and contact time of 5 s per step. High-angle XRD patterns were collected in the 2$\theta$ range between 20° and 100° with a step size of 0.02 s and contact time of 1 s per step.

Transmission Electron Microscopy (TEM) and energy dispersive X-ray spectroscopy (EDX) were carried out with a JEOL JEM 1400 or 2100 instruments operated at 120 and 200 kV, respectively (JEOL Ltd., Tokyo, Japan). Sample preparation was performed by dispersing *ca.* 1 mg of sample in 1 mL of 1-butanol followed by sonication in a low power sonicator bath (Selecta, Spain) for 5 min, and then depositing one drop of the suspension onto carbon-coated copper grids.

Textural properties of the materials were determined by $N_2$ adsorption porosimetry by using a Micromeritics ASAP 2020 (Micromeritics Co., Norcross, USA). To perform the $N_2$ measurements, *ca.* 30 mg of each sample was previously degassed under vacuum for 24 h at 40 °C. The surface area ($S_{BET}$) was determined using the Brunauer-Emmett-Teller (BET) method and the total pore volume ($V_t$) and pore volume ($V_p$) were estimated from the amount of $N_2$ adsorbed at relative pressures of 0.97 and 0.60, respectively. The pore size distribution between 0.5 and 40 nm was calculated from the desorption branch of the isotherm by means of the Barrett-Joyner-Halenda (BJH) method and the average mesopore size ($D_p$) was determined from the maximum of the pore size distribution curve.

Electrophoretic mobility measurements for the materials suspended in water were used to calculate the zeta-potential ($\zeta$) values of the nanosystems. Measurements were performed in a Zetasizer Nano ZS (Malvem Instruments Ltd., United Kingdom) equipped with a 633 nm "red" laser. For this purpose, 1 mg of nanoparticles was added to 10 mL of water followed by vortex and ultrasound to get a homogeneous suspension. Measurements were recorded by placing *ca.* 1 mL of the suspension in a DTS1070 disposable folded capillary cells (Malvern Instruments). Dilutions of the initial suspension were performed if needed. The hydrodynamic size of the nanoparticles was measured by dynamic light scattering (DLS) with the same Malvern instrument. Values presented are mean ± SD from quintuplicate measurements.

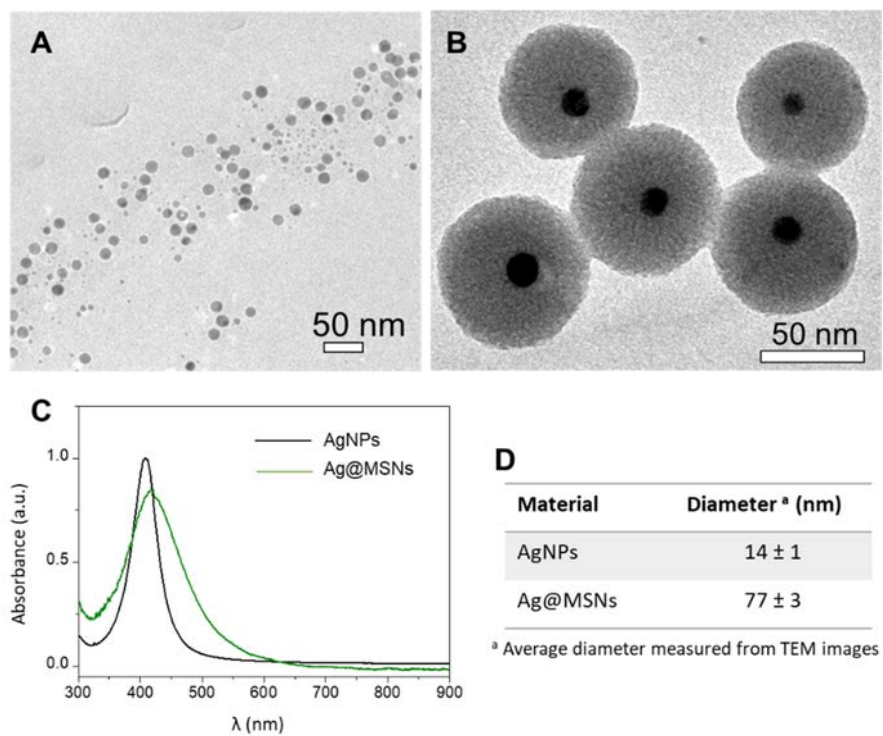

**Figure S1.** TEM micrographs of (A) silver nanoparticles and (B) surfactant containing Ag@MSNs. (C) UV-visible spectra of AgNPs in water and Ag@MSNs in ethanol. (D) Particle diameter obtained by measuring the nanoparticles in the TEM micrographs. Average diameters were calculated from 100 nanoparticles.

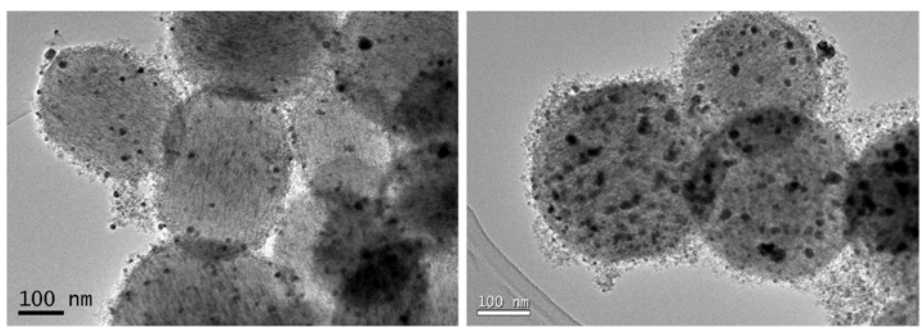

**Figure S2.** TEM micrographs of MSNs-AgBrNPs nanosystem.

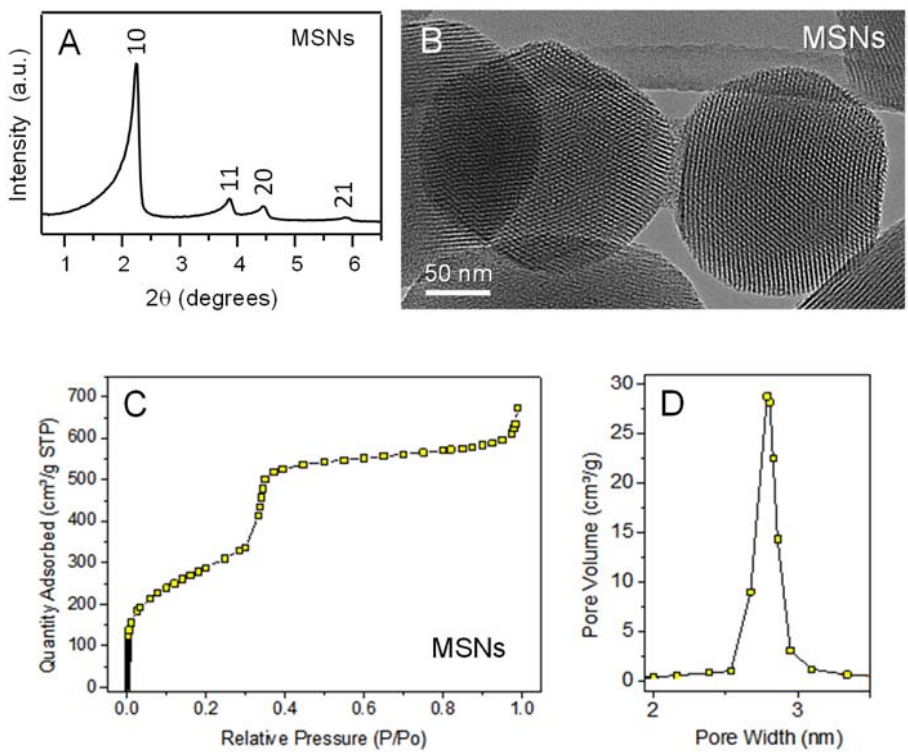

**Figure S3.** Characterization data of unmodified MSNs. A) Low angle powder X-ray diffraction pattern, B) TEM micrograph, C) $N_2$ adsorption-desorption isotherms and D) pore-size distribution. The textural parameters of the MSN materials obtained by $N_2$ sorption measurements are as follows: $S_{BET}$ = 1054.9 m$^2$/g, $D_P$ = 2.80 nm, $V_T$ = 0.946 cm$^3$/g and $V_P$ = 0.854 cm$^3$/g.

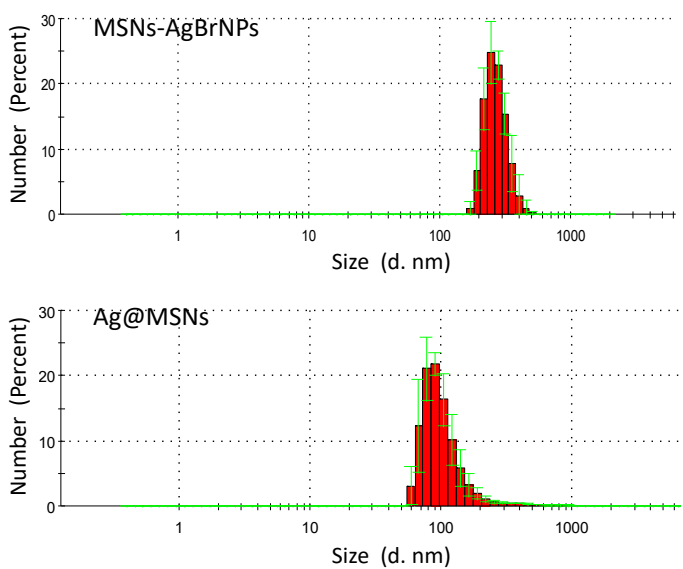

**Figure S4.** Hydrodynamic size distributions of materials measured by dynamic light scattering.

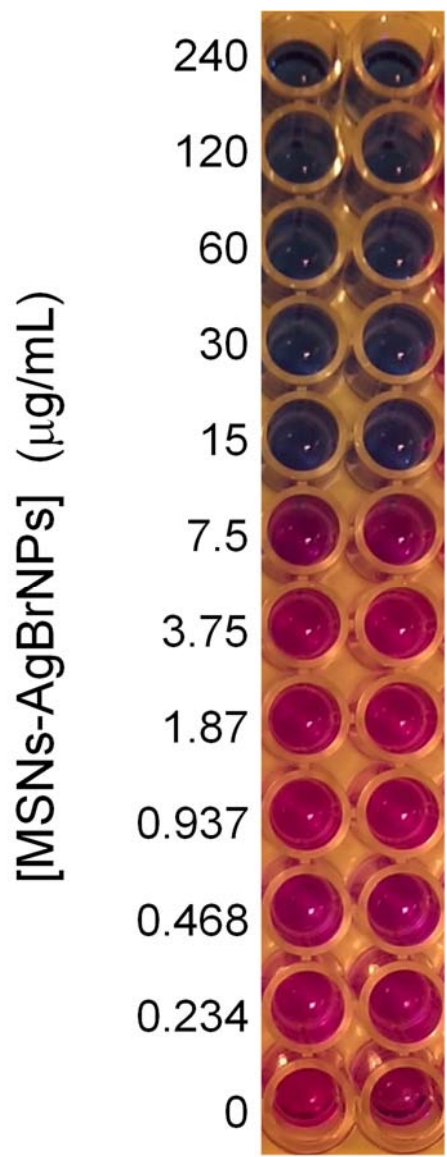

**Figure S5**. Determination of minimum inhibitory concentration (MIC) of MSN-AgBrNPs against *M. tuberculosis* using the resazurin test.

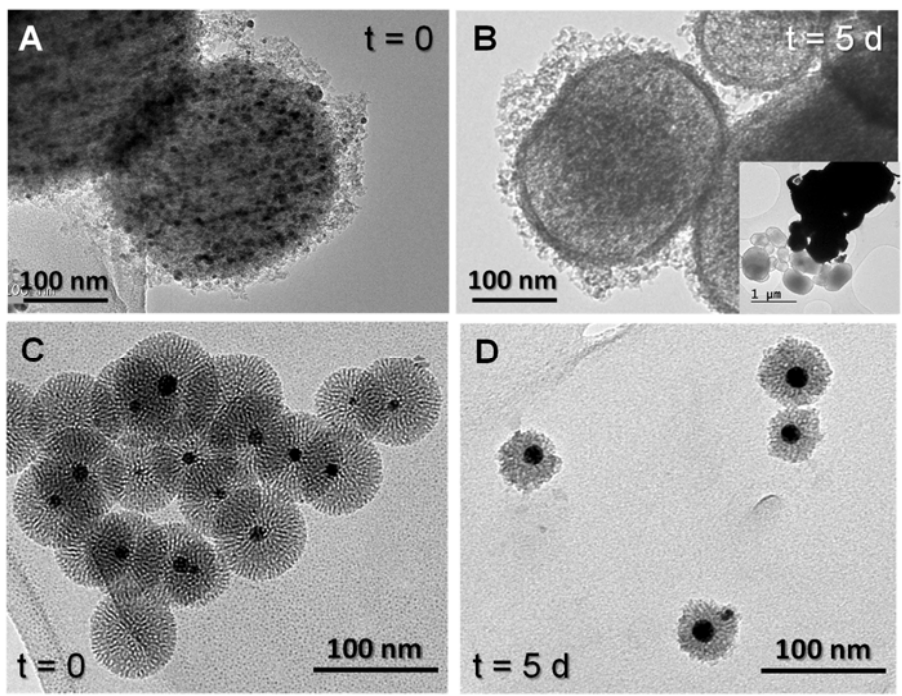

**Figure S6.** TEM images of MSNs-AgBrNPs (A, B) and Ag@MSNs (C, D) nanosystems before (A, C) and after 5 days of immersion in the *Mtb* culture medium (B, D).

**Table S1.** Energy-dispersive X-ray spectroscopy analysis of materials. Atomic percentages of Ag, Si and Br for MSNs-AgBrNPs and Ag@MSNs materials before and after 5 days of immersion in the *Mtb* culture medium.

| Material | Si K | Ag K | Br K | Ag/Si [a] |
|---|---|---|---|---|
| MSNs-AgBrNPs | 95.33 | 3.02 | 1.65 | 0.0317 |
| Ag@MS | 96.46 | 3.30 | 0.24 | 0.0342 |
| MSNs-AgBrNPs, t = 5 d | 76.69 | 16.28 | 7.03 | 0.2123 |
| Ag@MS, t = 5 d | 85.34 | 14.66 | --- | 0.1718 |

[a] silver to silicon molar ratio.

**Table S2.** Cumulative silicon concentration measured in the *Mtb* culture medium by ICP-AES for MSNs-AgBrNPs and Ag@MSNs materials. Values are mean of tree independent measurements.

| Time (days) | [Si] (mg/L) | |
|---|---|---|
| | MSN-AgBrNPs | Ag@MSNs |
| 0.25 | 47.5 | 23.4 |
| 1 | 95.0 | 54.9 |
| 2 | 141.5 | 92.8 |
| 3 | 186.8 | 125.1 |
| 5 | 232.1 | 168.6 |
| 7 | 262.3 | 191.8 |
| 14 | 291.4 | 216.9 |
| 21 | 319.8 | 239.8 |
| 28 | 345.8 | 260.8 |